# DESIGN, MANUFACTURING, ASSEMBLY, TESTING, AND LESSONS LEARNED OF THE PROTOTYPE 650 MHz COUPLERS *

J. Helsper[†], S. Chandrasekaran, N. Solyak, S. Kazakov, K. Premo, G. Wu,
F. Furuta, J. Ozelis, B. Hanna, FNAL, Batavia, IL 60510, USA

## Abstract

Six 650 MHz high-power couplers will be integrated into the prototype High Beta 650 MHz (HB650) cryomodule for the PIP-II project at Fermilab. The design of the coupler is described, including design optimizations from the previous generation. This paper then describes the coupler life-cycle, including manufacturing, assembly, testing, conditioning, and the lessons learned at each stage.

## INTRODUCTION

The prototype High Beta 650 MHz (pHB650) couplers will provide radio frequency (RF) input to the superconducting accelerating cavities housed within the pHB650 cryomodule (CM) [1], which is part of the PIP-II Project [2]. Six pHB650 couplers are used in the pHB650 CM string. Eight pHB650 couplers with three additional vacuum sides were procured. These pHB650 couplers are predated by 'proof of concept' 650 couplers, which validated the overall design and testing regime [3].

## DESIGN

The critical design components of the pHB650 coupler design are shown in Fig.1.

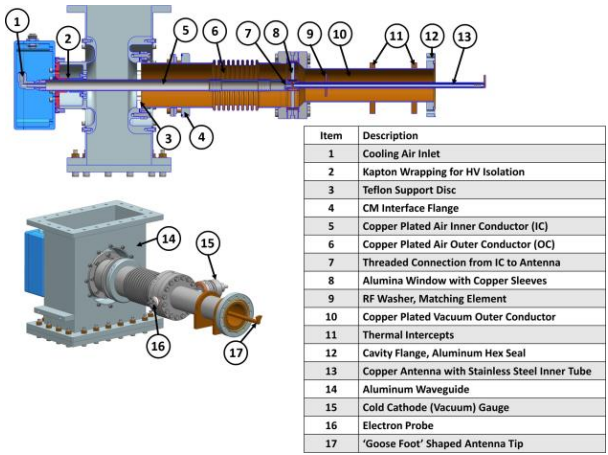

Figure 1: The full pHB650 Coupler Assembly

Brazed copper sleeves connect the window (Item 8) to its surrounding components, allowing for thermal expansion without undue stress. The stainless steel (SS) tube within the antenna provides an air cooling path and the stiffness necessary for transportation and handling. Ti-N coating was not applied to the Alumina window as a high voltage (HV) bias of 5 kV suppresses multipacting. The antenna tip (Item 17) has a non-symmetric 'goose foot' shape which allows for modulation of cavity Qext. The antenna assembly (Item 8+13), cold outer conductor (OC) (Item 10), air inner conductor (IC) (Item 5), and air OC (Item 6) were all designed to be vacuum furnace brazed.

### Changes from Previous Design

The two proof of concept coupler designs [3] [4] are shown in Fig.2. Design A used solid copper electromagnetic shielding, while design B used copper plating over the SS outer conductor.

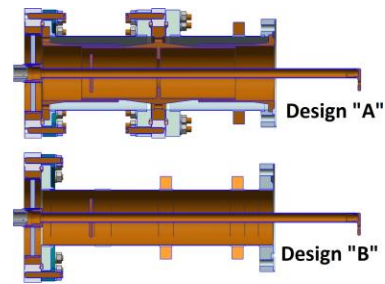

Figure 2: Section View, Proof of Concept Couplers

Design B was chosen as the basis for the pHB60 couplers as it was less complex, easier to assemble, and better for UHV cleaning. Several features of Design B were changed for the pHB650 coupler, which included a larger waveguide, a different electron probe location, inclusion of a vacuum gauge, increased OC wall thickness, a fully brazed SS tube within the antenna (previously removable design left the antenna easily deformed), and thru hole mounting on thermal straps instead of threaded inserts.

### Analysis

The RF and Thermal analyses performed for the pHB650 coupler [5] are similar in methods and results to those for the proof of concept couplers [4] [6], which show the couplers meet all necessary requirements. Previous structural analysis verified the thermal stress of the ceramic window sleeves [5], and recent analysis verified the coupler will have acceptable resonant frequencies and stress levels during shipment of the CM [7].

## MANUFACTURING

The manufacturing life cycle shown above in Fig. 3 was driven by Fermilabs's technical and procurement specifica-tions. The entire procurement and manufacturing of the couplers took place during the COVID-19 pandemic with travel restrictions, and as such, no Fermilab staff were allowed to visit the vendor. The main points of the manufacturing are summarized as follows.

* WORK SUPPORTED, IN PART, BY THE U.S. DEPARTMENT OF ENERGY, OFFICE OF SCIENCE, OFFICE OF HIGH ENERGY PHYSICS, UNDER U.S. DOE CONTRACT NO. DE-AC02-07CH11359.
[†] jhelsper@fnal.gov

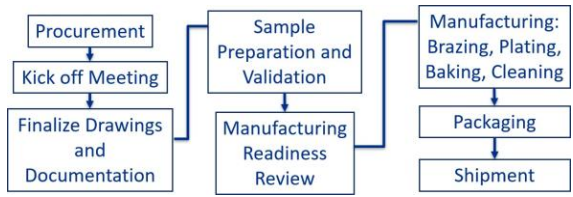

Figure 3: Manufacturing Life Cycle

Prior to manufacturing release the vendor created high quality brazed samples and redesigned all brazed joints in both tolerance and form based on their brazing experience. Upon manufacturing, several issues occurred, which highlighted the vendor's inexperience manufacturing couplers, which was in contrast to Fermilab's understanding during the procurement phase. This led to frequent involvement and oversight from Fermilab staff, and led to issues during manufacturing, such as damage to sealing surfaces without traceability, severe oxidation of IC at bellows during 400 °C vacuum treatment due a fast temperature ramp and poor heat conduction through bellows, sub-optimal electropolishing settings, oxidation on the outer conductor due to performing LN2 cold shock after 400 °C vacuum treatment, and delays receiving parts from subcontractors. Additionally, the vendor cleanroom was not to Fermilab requirements, and while the couplers passed 120 °C baking and RGA Scan criteria, couplers were cleaned and re-baked at Fermilab. The schedule was impacted due to several of these issues.

Lessons learned from manufacturing primarily revolve around the need for improved detail within the technical and procurement specifications. All issues that occurred during manufacturing, even minor, were documented and incorporated into the new specification. This experience reiterated the need for in-person vendor visits during fabrication, which could not be performed on this procurement due to the COVID-19 pandemic. Design improvements based on manufacturing include the need changing the copper antenna shaft to be made from solid stock instead of a tube (this likely contributed to the poor surface finish), and the mandatory redesign of brazed joints per vendor best practices using shoulder fits.

## INCOMING INSPECTION, CLEANING, ASSEMBLY, AND BAKING

Incoming inspection took place in an ISO 6 cleanroom and was followed travelers. Inspection of the couplers revealed the antenna surface was not mirror-like and was instead striated (as seen in Fig.4), there were often scratches on the antenna from metal-metal contact, many sealing surfaces were damaged, the units were not UHV clean, and initial units of the vacuum OC showed heavily oxidized plating (as seen in Fig.5. All antennas appeared to 'bend' along their length, but only a few were found to have displacement at the antenna tip. This observation was later confirmed with a coordinate measurement machine on one antenna, which was found to have the tip position within tolerance while other locations were not. All of these non-conformities were accepted by Fermilab, and all corrected except for the antenna straightness, which was deemed acceptable. Other than these issues, the craftsmanship of the couplers was excellent, particularly the braze quality, as seen in Fig.4. Cleaning of the couplers was routine.

Assembly to the RF chamber only saw minor challenges, such as the 5 K intercept interfering with proper stud installation. The tooling used for installation, while needing some improvements for stability and ease of movement, served its purpose. Fig.6 shows the assembly to the RF chamber in process. The post bake RGA requirements were easily surpassed during the 48 hour 120 °C bake and subsequent RGA scan, and the RGA results are shown in Fig.7.

Lessons learned for this phase include the need for thermal intercept optimization, that the antenna design must be improved since the bending was found to be caused by buckling which occurs upon contraction after the 800 °C braze, and that dedicated inspection tooling should be created for production couplers as manually separating the clean antenna and OC are a non-trivial procedure.

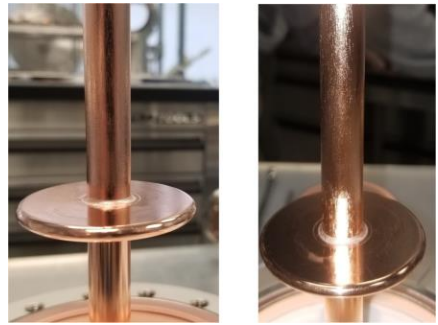

Figure 4: Antenna Braze Quality and Striations. Left: Normal Lighting, Right: With Flash

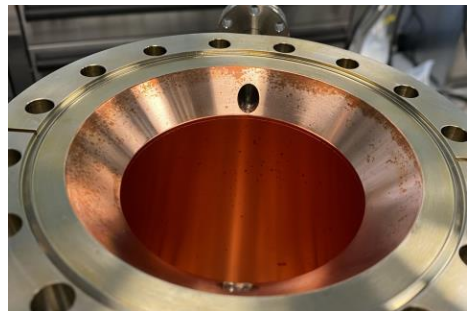

Figure 5: Plating Oxidation on Vacuum Outer Conductor

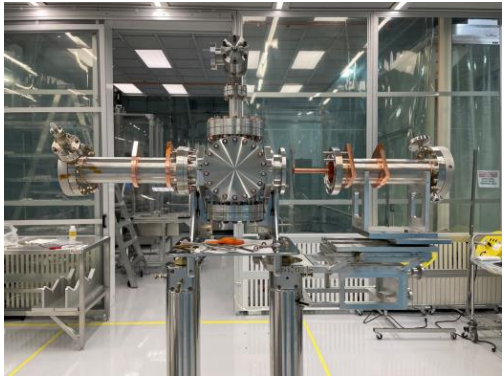

Figure 6: Coupler Assembly to RF Test Chamber

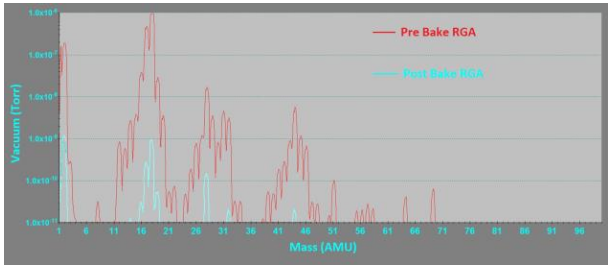

Figure 7: RGA Scan Results Before and After 120 °C Baking

## WARM RF TESTING AND RESULTS

All assemblies were conditioned, with the exception of one vacuum side assembly. The RF test stand is shown in Fig.8. Water cooled thermal straps attached to the ceramic window flange are used to provide cooling. Temperature sensors on the ceramic window flange, air outlet, water outlet, and measurements of vacuum and DC bias were the RF interlocks.

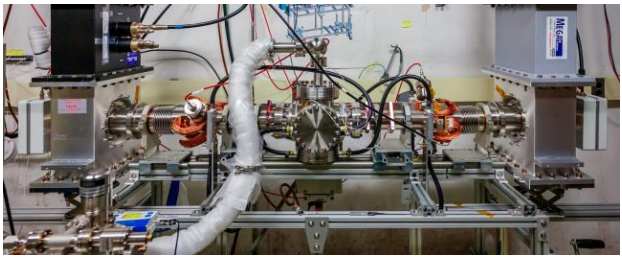

Figure 8: RF Test Stand

The test stand reached approx. 100 kW in standing wave operation. Each pair of couplers was tested at 8 reflection phases. The couplers were conditioned to reach 50 kW at full reflection. Conditioning started with 1 Hz pulses of 10 ms, which proceeded to 20, 50, 100, 200, and 500 ms, with power being ramped from 0 to 100 kW. Power and pulse duration did not increase unless vacuum levels were below 2e-6 Torr. Then power was ramped to 50 kW in continuous wave mode (CW) and maintained until ceramic window's temperature saturated. Final acceptance required the window flange to be <60 °C and vacuum < 1e-7 Torr, which was achieved for all couplers tested. RF power and vacuum activity plots are shown in Fig.9 for coupler pair 8+9. It shows that multipacting occurred in the first two phases, but afterwards subsided.

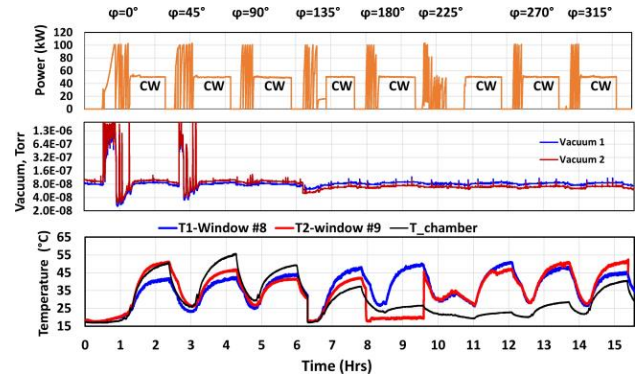

Figure 9: Couplers 8+9 RF Power, Vacuum and Temperature History

## SUMMARY

In summary, even though COVID prevented any visitation from Fermilab to the vendor, all pHB650 couplers tested have been successfully qualified for use on cavities without major incident. Procurement, manufacturing, QC, cleaning, assembly, testing, and integration have all provided valuable experiences which will positively influence the production 650 MHz coupler design and specifications, along with those of the other PIP-II couplers at Fermilab.

## ACKNOWLEDGEMENTS

The work completed would be impossible without Fermilab's dedicated technicians D. Plant, R. Kirschbaum, M. Chlebek, M. Battistoni, B. Tennis, and their supervisor D. Bice.

## REFERENCES

[1] V. Roger, N. Bazin, S.K. Chandrasekaran, S. Cheban, M. Chen, R. Cubizolles, *et al.*, "Design of the 650 MHz High Beta Prototype Cryomodule for PIP-II at Fermilab", in *Proc. SRF'21*, East Lansing, MI, USA, Jun.-Jul. 2021, pp. 671–676. doi:10.18429/JACoW-SRF2021-WEPTEV015

[2] A.L. Klebaner, C. Boffo, S.K. Chandrasekaran, D. Passarelli, and G. Wu, "Proton Improvement Plan ' II: Overview of Progress in the Construction", in *Proc. SRF'21*, East Lansing, MI, USA, Jun.-Jul. 2021, pp. 182–189. doi:10.18429/JACoW-SRF2021-MOOFAV05

[3] S. Kazakov, B.M. Hanna, O.V. Pronitchev, and N. Solyak, "Latest Progress in Designs and Testings of PIP-II Power Couplers", in *Proc. SRF'19*, Dresden, Germany, Jun.-Jul. 2019, pp. 263–266. doi:10.18429/JACoW-SRF2019-MOP080

[4] S. Kazakov, "Couplers Activities in FNAL", presented at *Worldwide Workshop in Fundamental Power Couplers*, CERN, Jun, 2019. https://indico.cern.ch/event/827811/

[5] S. Kazakov, "RF Couplers for PIP-II", presented at *PIP-II 650 MHz Coupler proto-Final Design Review*, Fermilab, Feb, 2020. https://indico.fnal.gov/event/22769/

[6] S. Kazakov, "RF Couplers for PIP-II", presented at *Worldwide Workshop in Fundamental Power Couplers*, CERN, Jun, 2017. https://indico.cern.ch/event/642503/contributions/

[7] J. Helsper, S. Cheban, and I. Salehinia, "Transportation Analysis of the Fermilab High-Beta 650 MHz Cryomodule", in *Proc. SRF'21*, East Lansing, MI, USA, Jun.-Jul. 2021, pp. 682–686. doi:10.18429/JACoW-SRF2021-WEPTEV017